\documentclass[twocolumn,showpacs,preprintnumbers,amsmath,amssymb,10pt]{revtex4-1}
\usepackage{graphicx}
\usepackage{dcolumn}
\usepackage{bm}
\usepackage{natbib}
\begin{document}

\title{Pulses of chaos synchronization in coupled map chains with delayed transmission}

\author{Bernhard Schmitzer}
\affiliation{Institute for Theoretical Physics, University of W\"urzburg, 97074
W\"urzburg, Germany}
\author{Wolfgang Kinzel}
\affiliation{Institute for Theoretical Physics, University of W\"urzburg, 97074
W\"urzburg, Germany}
\author{Ido Kanter}
\affiliation{Department of Physics, Bar-Ilan University, Ramat-Gan, 52900 Israel}

\date{\today}

\begin{abstract}
Pulses of synchronization in chaotic coupled map lattices are discussed in the context of transmission of information. Synchronization and desynchronization propagate along the chain with different velocities which are calculated analytically from the spectrum of convective Lyapunov exponents. Since the front of synchronization travels slower than the front of desynchronization, the maximal possible chain length for which information can be transmitted by modulating the first unit of the chain is bounded.
\end{abstract}
\pacs{05.45.Ra, 05.45.Vx, 05.45.Xt}
\maketitle
\section{Introduction}
Chaos synchronization is a counter-intuitive phenomenon which has been extensively investigated since its original discovery \cite{Pecora:1990lang,Pikovsky1984}. In particular, its potential of being applied for novel secure communication devices has attracted a lot of research on the foundations of chaos synchronization \cite{Pikovsky:Buchlang,BoccalettiKurthslang,BoccalettyLatoralang}. Motivated by experiments on synchronized chaotic lasers, networks of chaotic units with time-delayed couplings and feedback are recently being studied \cite{Arenas,UchidaRogister}. For chains of chaotic units with delayed couplings other counter-intuitive phenomena have been found such as anticipated or sublattice synchronization \cite{Voss,KestlerKinzelKanter}. Chains of chaotic units are also discussed in the context of convective instabilities.
Instabilities of such systems have successfully been described by spectra of comoving Lyapunov exponents
\cite{PhysRevLett.85.3616,BohrRand1991,Pikovsky1993,AransonGolombSompolinsky1992,RudzickPikovsky1996,MendozaBoccalettiPoliti2004}.

There are several possibilities to transmit a message between two synchronized chaotic units A and B \cite{KinzelHandbooklang}. One possibility is chaos modulation: the message
 modulates the dynamics of unit A. This affects the dynamics of unit B and it is possible to recover the message with a certain bit error rate \cite{Cuomo:1993lang,Kocarev:1995lang}.
This application is the motivation for the following question which we investigated in this report: is communication based on chaos modulation possible in a long chain of unidirectional coupled chaotic units?

\begin{figure}
\includegraphics{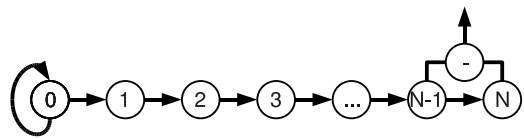}
\caption{Scheme of the examined chain: each unit is driven by its predecessor.}
\label{fig:scheme}
\end{figure}

The dynamics of the first unit is changed according to a given bit sequence. These bits are recovered by observing the synchronization of the two last units of the chain. Note that we keep the system simple: the bits are not recovered for each unit but only with the two last units of the chain indicated in Fig. \ref{fig:scheme}. The intermediate units act as passive relay stations which may be necessary to transmit signals over long distances.
Note that the units can transmit the signals with arbitrary delay times. In order to detect perturbations of the first unit at the end of the chain we introduce a self-feedback of the first sending unit with a delay time which is identical to the coupling delay of the last two units. We analyze the transmission of information using the method of comoving Lyapunov exponents.

In fact already for single units with delayed feedback one finds, in a spatiotemporal representation, convective instabilities which move in a cone with a spectrum of velocities \cite{PhysRevLett.76.2686}. A chain of oscillators, nonlinear partial differential equations close to instabilities, and lasers with delayed feedback have been analyzed with the spectrum of comoving Lyapunov exponents \cite{PhysRevLett.85.3616}. Chains of chaotic units which have been discussed in the context of turbulence show convective instabilities as well \cite{BohrRand1991,Pikovsky1993,AransonGolombSompolinsky1992,RudzickPikovsky1996}, even if the single units have a delayed self-feedback which generates anticipated chaos \cite{MendozaBoccalettiPoliti2004}.

In this report we extend this analysis to time-delayed couplings in the context of secure communication.
\section{System dynamics}
In this report we investigate the previous general question for a simple model: a chain of coupled maps with time-delayed transmission. The first unit
 has a self-feedback which is necessary to obtain zero lag synchronization for the last two units of the chain. The system is defined by the following equations:
\begin{eqnarray}
x_t^0 = (1-\varepsilon)\ f(x_{t-1}^0) + \varepsilon\ f(x_{t-\tau_0}^0) \nonumber \\
x_t^n = (1-\varepsilon)\ f(x_{t-1}^n) + \varepsilon\ f(x_{t-\tau_n}^{n-1})
\label{eqn:general_update}
\end{eqnarray}
$x_t^n$ is the variable of the $n$th unit at the discrete time step $t$. $f(x)$ is a map which yields chaotic iterations. The transmission time from unit $n-1$ to unit $n$ is denoted by $\tau_n$ and $\varepsilon$ is the coupling strength. The first unit $n=0$ has self-feedback with delay $\tau_0$.
It is easy to see that the system has the following synchronized solution:
\begin{eqnarray}
x_t^n=x_{t-\Delta_n}^0 \nonumber \\
\Delta_n=\sum_{k=1}^n \tau_k - n \cdot \tau_0
\label{eqn:general_solution}
\end{eqnarray}
If all delays are identical to the feedback delay of the first unit, $\tau_n= \tau_0$, one finds complete zero lag synchronization, $x_t^n=x_t^0$. For smaller values of $\tau_n$ one finds anticipated chaos whereas for larger ones the units lag behind the first one. When the transmission delay from unit $N-1$ to unit $N$ is identical to the feedback $\tau_0$ of the first unit, the last two units are synchronized without time shift, $x_t^{N-1}=x_t^N$. Hence, the synchronization of the chain can be measured from the difference of the variables of the last two units.
From now on identical delay times $\tau_n=\tau_0$ are used, since different $\tau_n$ lead to trivial time shifts.
The stability of the chaotic synchronized solution $x^n_t=x^0_t$ can be calculated by considering small perturbations of this trajectory and linearizing Eq. (\ref{eqn:general_update}). The evolution of deviations between two consecutive units, $\delta^n_t=x^n_t-x_t^{n-1},\ n=1,\ldots,N$ are determined by the following set of linear equations:
\begin{eqnarray}
\delta_t^1 = (1-\varepsilon)\ f'(x_{t-1}^0) \delta_{t-1}^1 \nonumber \\
\delta_t^n = (1-\varepsilon)\ f'(x_{t-1}^0) \delta_{t-1}^n + \varepsilon\ f'(x_{t-\tau_0}^0) \delta_{t-\tau_0}^{n-1}
\label{eqn:delta_update}
\end{eqnarray}
If only the unit $n$ is perturbed, $\delta^k_t=0$ for $k<n$, then the deviation $\delta^n_t$ follows the equation
\begin{equation}
\delta_t^n = (1-\varepsilon)\ f'(x_{t-1}^0) \delta_{t-1}^n
\end{equation}
This perturbation is stable if
\begin{equation}
\langle \ln|f'(x_t^0)| \rangle < - \ln(1-\varepsilon)
\label{eqn:lyapunov_condition}
\end{equation}
Note that the average $\langle \ldots \rangle$ is taken over the first unit which, due to the feedback delay $\tau_0$, is hyperchaotic; i.e., it has a spectrum of positive Lyapunov exponents. According to Eq. (\ref{eqn:lyapunov_condition}), the synchronized solution (\ref{eqn:general_solution}) is stable if the coupling strength $\varepsilon$ is large enough.
\section{Chaos Modulation}
Now we apply chaos modulation to a concrete system: we take $f(x)=r\,x\,(1-x)$ with $r=4$ for the bulk of the chain and a variable $r_0$ for the first unit depending on the message to transfer. If $r_0=r=4$ then the chain will successively relax into the synchronized state starting from the head. If one chooses $r_0=r'\not = r$, for example, $r'=3.95$, the system leaves the common trajectory, again beginning at the head. This creates pulses of (de)synchronized phases traveling throughout the chain. We will call $r_0=r$ the tuned and $r_0\not = r$ the untuned state.
\begin{figure}
\includegraphics{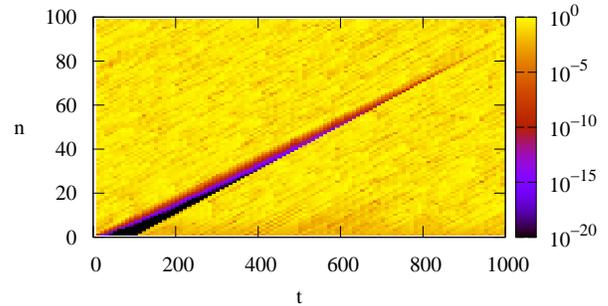}
\caption{Absolute value of the difference $\delta^n_t$ between neighboring units: for $t<0$, $100<t$ the first unit is untuned ($r_0=r'=3.95$). For $0<t<100$ the chain is tuned ($r_0=r=4$), thus, creating a synchronized pulse. $\varepsilon=0.8, \tau_0=10$.}
\label{fig:pulse}
\end{figure}
Figure \ref{fig:pulse} shows numerical results of a simulation for $\varepsilon=0.8,\tau_0=10$: the chain starts in a desynchronized state and is tuned at $t=0$. After 100 time steps it is untuned again. The absolute difference $| \delta_t^n | = |x_t^n-x_t^{n-1}|$ is shown in color code; i.e., the dark areas mark synchronization.

The velocities of the signal fronts are defined as
$$v=\frac{\Delta \mbox{number of units}}{\Delta \mbox{time}}$$
In numerical simulations two units are considered synchronized if $|\delta_t^n|<\Theta$ with a small threshold $\Theta\approx10^{-10}$. We find that for $\Theta \ll 1$ the results for the velocities of the head and tail of the pulse do not depend on the actual choice of the threshold.
In the simulation shown in Fig. \ref{fig:pulse} one finds that synchronization travels with a speed of approx $v_s=0.088$ whereas desynchronization propagates with $v_d=0.100$ which is the maximum possible velocity defined by the coupling delay $\tau_0=10$. Thus, desynchronization finally overtakes synchronization and the pulse cannot be detected in the rest of the chain.

\begin{figure}
\includegraphics{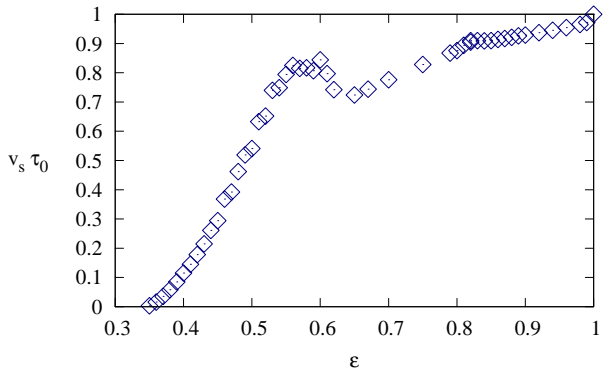}
\caption{Propagation velocity of synchronization in units of the maximum velocity $1/\tau_0$ for different coupling parameters $\varepsilon$ at $\tau_0=10$. Here the critical coupling is $\varepsilon_c\approx0.35$.}
\label{fig:vel_eps}
\end{figure}

Figure \ref{fig:vel_eps} shows $v_s$ as a function of the coupling strength $\varepsilon$. Near the critical coupling one observes a slowing down. One would expect a monotonous decrease in the speed with decreasing coupling strength. However, there are two small peaks in the data. For $\tau_0>1$ the value distribution of the first unit and thus the local Lyapunov exponent $\lambda_0=\langle \ln |f'| \rangle$ averaged over this distribution are functions of $\varepsilon$. Near the peaks $\lambda_0$ becomes rather small making the map ``less chaotic'' or even non-chaotic and thus accelerating synchronization.

\begin{figure}
\includegraphics{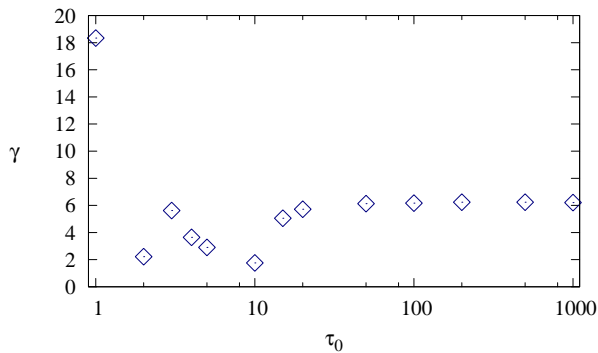}
\caption{The additional delay $\gamma=1/v_s-\tau_0$ as a function of the coupling delay $\tau_0$ with $\varepsilon=0.6$: for large $\tau_0$ there is a saturation effect.}
\label{fig:vel_tau}
\end{figure}

Figure \ref{fig:vel_tau} shows the additional delay per unit $\gamma$ with $v=1/(\tau_0+\gamma)$ as a function of $\tau_0$. For large $\tau_0$ there is a saturation effect.

\section{Convective Perturbations}
For a chain of connected relay stations one might expect the coupling delay between the units to be on a much larger time scale than the one of the internal dynamics. In our model this would imply $\tau_0 \gg 1$. However the numerical results presented by now indicate that the value of $\tau_0$ has a quantitative effect on the velocities but does not change the behavior of the chain in principle (see Fig. \ref{fig:vel_tau}). Hence, for analytical treatment we shall focus on $\tau_0=1$.
We start with a completely synchronized chain, $x^n_t=x^0_t$, and consider small perturbations, Eq. (\ref{eqn:delta_update}).
Then the discrete Green's function of the system is \cite{Pikovsky1993}
\begin{eqnarray}
\label{eqn:Greens_function}
g^n_t=
\left(
\begin{array}{c}
	t \\
	n
\end{array}
\right)
\,
(1-\varepsilon)^{t-n}\,\varepsilon^n\,\prod_{i=1}^t f'(x^0_{i-1})
\end{eqnarray}
The evolution of an arbitrary perturbation $\delta^k_0$ of the synchronized solution can be obtained by
$$
\delta^n_t=\sum_k \delta_0^k\ g_t^{n-k}
$$
The Green's function can be approximated by
\begin{eqnarray}
\left|g^n_t\right|\approx\exp(\Lambda(v)\,t)\,,~v=n/t\\
\label{eqn:convective_exponent}
\Lambda(v)=(1-v)\,\ln\left(\frac{1-\varepsilon}{1-v}\right)+v\,\ln\left(\frac{\varepsilon}{v}\right)+\lambda_0
\end{eqnarray}
where $\lambda_0=\langle \ln |f'(x^0_t)| \rangle$ is the Lyapunov exponent of the applied map and $\Lambda(v)$ is the convective Lyapunov exponent as mentioned in \cite{BohrRand1991,Pikovsky1993,AransonGolombSompolinsky1992,RudzickPikovsky1996,MendozaBoccalettiPoliti2004}. Outside the interval $v\in [0:1]$ $\Lambda(v)$ has to be set $-\infty$ due to causality.
\begin{figure}
\includegraphics{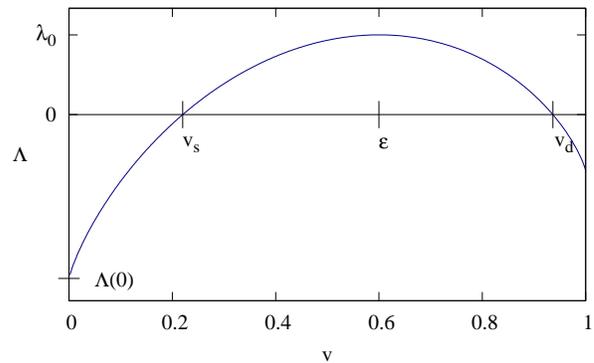}
\caption{The convective Lyapunov exponent $\Lambda(v)$, Eq. (\ref{eqn:convective_exponent}), for $\lambda_0=0.3$ and $\varepsilon=0.6$.}
\label{fig:convective_lambda}
\end{figure}
Figure \ref{fig:convective_lambda} shows the convective Lyapunov exponent $\Lambda(v)$ of Eq. (\ref{eqn:convective_exponent}). $\Lambda(v\rightarrow 0)$ describes the relaxation of a local perturbation, for negative values the chain finally relaxes to the synchronized trajectory, in agreement with Eq. (\ref{eqn:lyapunov_condition}). The fronts of desyncronization and resynchronization are at the border of stability, $\Lambda(v)=0$. Hence, the first zero at $v_s$ determines the velocity of synchronization and the second one yields the velocity of desynchronization. If $\ln(\varepsilon)>-\lambda_0$ one finds the maximal possible desynchronization speed $v_d=1$.

Hence, a local perturbation of the synchronized trajectory generates a pulse of deviations with a front velocity $v_d$ and a tail velocity $v_s$. The maximum of the pulse propagates with $v_m=\varepsilon$ and the integrated pulse increases exponentially with the exponent $\lambda_0$.

\begin{figure}
\includegraphics{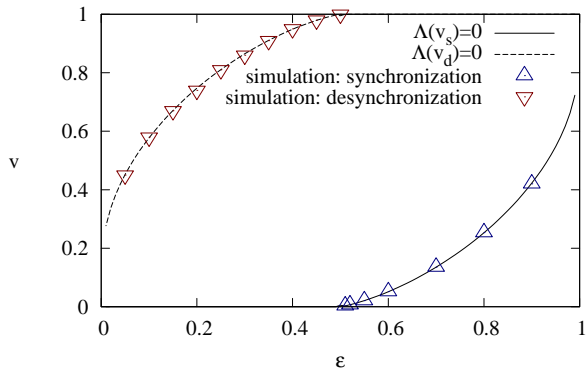}
\caption{$v_s$ and $v_d$ obtained by simulations are compared to the predictions made by $\Lambda(v_i)=0$ for $\tau_0=1$.}
\label{fig:Lambda_prediction}
\end{figure}

Figure \ref{fig:Lambda_prediction} compares this theory with results obtained from simulations. Obviously the prediction of $\Lambda(v)$ is correct even if the linear equations do not describe the dynamics in the noisy regions of the chain preceding the final synchronization process.
Note that for $\tau_0=1$ one finds $\lambda_0=\ln2$; hence, the critical coupling is $\varepsilon_c=1/2$. At this critical point the velocity of synchronization goes to zero; again one observes a critical slowing down for synchronization.
\section{Information Transfer}
 Pulses of synchronization and desynchronization can be used for communication: sender and receiver define a certain bit length $L$. Then the sender switches $r_0$ between $r$ and $r'$ in time intervals $L$ according to the bit sequence (let us define $r_0=r \equiv$ true, $r_0=r' \equiv$ false). Regions of corresponding state are created at the head and start to propagate throughout the chain. The receiver decodes the message by observing the synchronization state of the last two units.

It is in any case $v_s<v_d$ so the true-bits will always suffer from shrinking and a single true-bit will finally vanish at a chain position approximately determined by
\begin{equation}
	n^\ast \approx L \cdot \frac{v_d\,v_s}{v_d-v_s}
	\label{eqn:bitlength}
\end{equation}
However if a true-bit is followed by a second one, the second bit can travel a longer distance before vanishing. Obviously bit sequences with consecutive ones such as 01110 can be transmitted with much lower bit error rates than isolated ones such as 01010.
\section{Conclusion}
Pulses of synchronization in chaotic chains can be used for communication. The information has not to be restored at every unit but is recovered by observing the synchronization status of the last two units. We have analyzed the propagation of pulses using the method of convective Lyapunov exponents \cite{PhysRevLett.85.3616,BohrRand1991,Pikovsky1993,AransonGolombSompolinsky1992,RudzickPikovsky1996,MendozaBoccalettiPoliti2004,PhysRevLett.76.2686}. We find that the front of a synchronization pulse travels slower than its tail.
 This limits the maximal chain length for which the pulse can be detected and, thus, has a large influence on the bit error rate.

We have investigated the propagation of binary information by detuning the first chaotic unit. For a chain of chaotic lasers this may be accomplished by detuning the pump current of the first laser. It would be interesting to investigate the conditions for which analog signals can be transmitted. From our results we expect two major criteria for reliable transmission: The signal must have a spectrum of low frequencies compared to the inverse minimal bit length (\ref{eqn:bitlength}) and the linear approximation of the time evolution must hold; see Eq. (\ref{eqn:delta_update}).

\section*{ACKNOWLEDGEMENT}
We would like to thank Arkady Pikovsky for useful comments.
\bibliography{CHAOS,CUSTOM}{}
\bibliographystyle{apsrev4-1}
\end{document}